\documentclass[final,1p,times]{elsarticle} 
\usepackage{graphicx} 
\usepackage{amssymb} 
\usepackage{amsthm} 
\usepackage{lineno} 

\journal{Nuclear Physics A} 
\begin{document} 

\begin{frontmatter} 


\title{Conical Correlations, Bragg Peaks, and Transverse Flow Deflections
in Jet Tomography}

\author{Barbara Betz$^{a}$}
\address[a]{Institut f\"ur Theoretische Physik, 
Johann Wolfgang Goethe-Universit\"at, Frankfurt am Main, Germany}
\author{Jorge Noronha$^{b}$}
\address[b]{Department of Physics, Columbia University, 
New York, 10027, USA}
\author{Giorgio Torrieri$^{a,c}$}
\address[c]{Frankfurt Institute for Advanced Studies (FIAS), 
Frankfurt am Main, Germany}
\author{Miklos Gyulassy$^{b}$}
\author{Dirk H.\ Rischke$^{a,c}$}

\begin{abstract} 
We use (3+1)-dimensional ideal hydrodynamics to describe a variety of different
jet energy loss scenarios for a jet propagating through an opaque medium. The
conical correlations obtained for fully stopped jets, revealing a Bragg
peak, are discussed as well as results from pQCD and AdS/CFT. Moreover, 
we investigate transverse flow deflection. It is demonstrated that a double-peaked
away-side structure can be formed due to the different contributions of several
possible jet trajectories through an expanding medium.

\end{abstract} 

\end{frontmatter} 



\section{Introduction}

The observation of a double-peaked structure in azimuthal di-hadron correlations 
\cite{2pc} arose a lot of recent interest, since it was suggested 
\cite{MachConeIdea} that this structure could be related to the emission angles of
Mach cones that are via Mach's law ($\cos\phi_M= c_s/v_{\rm jet}$) directly related 
to the Equation of State (EoS).
In general, energetic back-to-back jets produced in the early stages of a 
heavy-ion collision propagate through the medium, depositing energy
and momentum along their path. Certainly, the properties of this deposition
depends on the physics of the jet-medium interactions. 
Recently, different mechanisms of jet energy loss were analyzed, ranging from weak 
\cite{Neufeld} to strong coupling \cite{MachAdSCFT,Noronha}. While the static 
background offers the possibility to compare to results obtained from AdS/CFT, 
the expansion of a system formed in a heavy-ion collision will certainly 
influence any kind of jet deposition scenario \cite{Satarov:2005mv}. 
We demonstrate that taking into account various possible trajectories of jets
propagating through the plasma \cite{Chaudhuri}
lead to a double-peaked structure on the away-side of the final particle correlations.
Here, the medium is investigated using Glauber initial conditions, 
corresponding to a gold nucleus with $r=6.4$~fm and a maximum temperature of $T_{\rm max}=200$~MeV. 
We focus on radial flow only and consider most central collisions, thus neglecting any 
elliptic flow contribution. For simplicity, the medium is always considered as an ideal gas of 
massless SU(3) gluons.
In our system of coordinates, the beam axis is pointing into the $z$ direction and the
associated jet moves along the $x$ direction. 

\section{A Hydrodynamical Prescription of Jets}
\begin{figure}[t]
\centering
  \includegraphics[scale = 0.56]{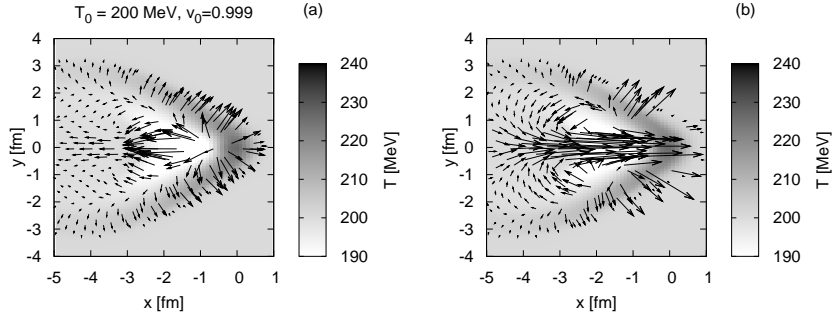}
  \caption{Temperature pattern and flow velocity profile (arrows) after a
  hydrodynamical evolution of $t=4.5$~fm for a jet that decelerates according to the
  Bethe--Bloch formula and stops after $\Delta x=4.5$~fm. The jet's initial velocity 
  is $v_{\rm{jet}}=0.999$. In the left panel a vanishing momentum loss rate is 
  assumed while in the right panel energy and momentum loss are considered according to the
  Bethe--Bloch formalism.}
  \label{Fig1}
\end{figure}
\begin{figure}[t]
\centering
  \includegraphics[scale = 0.56]{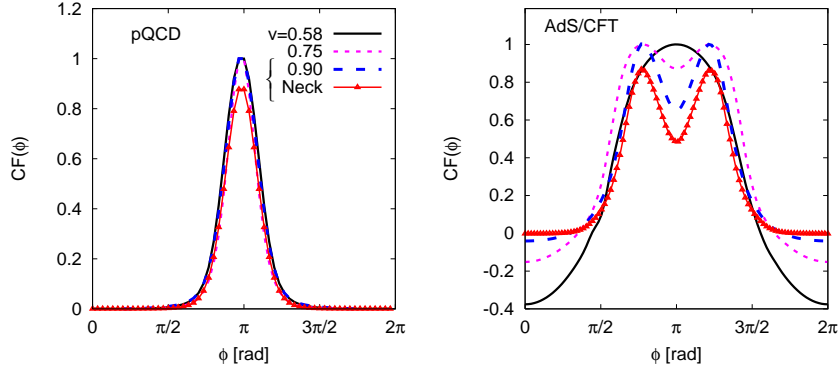}
  \caption{Normalized (and background subtracted) azimuthal away-side jet associated correlation 
  after Cooper-Frye freeze-out $CF(\phi)$ for pQCD {\protect\cite{Betz:2008wy}} (left) and AdS/CFT 
  from {\protect\cite{Noronha}} (right). Here $CF(\phi)$ is evaluated at 
  $p_T=5 \pi \,T_0 \sim 3.14$ GeV and $y=0$ for different jet velocities of
  $v=0.58, 0.75,0.9$. The line with triangles represents the Neck 
  contribution (which is a region close to the head of the jet) for a jet with $v=0.9$.}
  \label{CooperFrye_all}
\end{figure}

  
Assuming that the energy lost by the jet thermalizes quickly \cite{Adams:2005ph},
we solve the ideal hydrodynamical equations using a ($3+1$)-dimensional
SHASTA algorithm \cite{Rischke:1995pe}
\begin{eqnarray}
\partial_\mu T^{\mu\nu}&=&S^\nu\,.
\end{eqnarray}
The source term for the decelerating jet is given by
\begin{eqnarray}
\label{source}
S^\nu = \int\limits_{\tau_i}^{\tau_f}d\tau 
\frac{dM^\nu}{d\tau}\delta^{(4)}
\left[ x^\mu - x^\mu_{\rm jet}(\tau) \right],
\end{eqnarray}
where $\tau_{f}-\tau_i$ denotes the proper time interval 
associated with the jet evolution and $dM^\nu/d\tau = (dE/d\tau,d\vec{M}/d\tau)$ is
the energy and momentum loss along the trajectory of the jet
$x^\mu_{\rm jet} (\tau) = x_0^\mu + u^\mu_{\rm jet}\tau$. We assume that 
$dE(t)/dt=a/v_{\rm jet}(t)$, according to the Bethe--Bloch formalism \cite{Bragg}, 
leading to a Bragg peak as demonstrated in Ref.\ \cite{Betz:2008ka}.
For a jet starting at $v_{\rm jet}=0.999$, the initial energy loss rate can be determined
by imposing that the jet stops after $\Delta x=4.5$~fm, resulting in $a\simeq -1.3607$~GeV/fm 
\cite{Betz:2008ka}.
Fig.\ \ref{Fig1} displays the temperature and flow velocity profiles
of a jet with an energy loss as determined above and vanishing momentum
deposition (left panel) as well as an energy and momentum deposition (right panel). In
the latter case, the creation of a diffusion wake behind the jet is clearly visible, which 
leads to an away-side peak in the associated jet direction \cite{Betz:2008ka} after performing 
a Cooper--Frye (CF) \cite{Cooper:1974mv} freeze-out. 
Considering vanishing momentum deposition, the away-side peak is replaced by a conical 
(double azimuthal peak) distribution at the expected Mach cone angle \cite{Betz:2008ka}. 
\\\hspace*{4mm}
The away-side diffusion peak in the particle correlation also prevails when considering the Neufeld pQCD
source term \cite{Neufeld,Betz:2008wy}, because it involves a large 
momentum deposition (see Fig.\ \ref{CooperFrye_all}). The freeze-out results of the
AdS/CFT solution, however, show a double-peaked structure in spite of the diffusion plume due to a
novel nonequilibrium strong coupling effect in the ``Neck'' region as shown in Ref.\
\cite{Noronha}.
\\\hspace*{4mm}
For the expanding medium, we choose the following ansatz for the energy and momentum depostition 
of the jet, scaling with the temperature of the dynamical background
\begin{eqnarray}
\label{SourceExpandingMedium0}
S^\nu = \int\limits_{\tau_i}^{\tau_f}d\tau 
\frac{dM^\nu}{d\tau}\Big\vert_0\left[\frac{T(t,\vec{x})}{T_{\rm max}}\right]^3
\delta^{(4)} \left[ x^\mu - x^\mu_{\rm jet}(\tau) \right],
\end{eqnarray}
where $dE/dt_0 = 1$~GeV/fm and $dM/dt_0 = 1/v dE/dt_0$. Since deceleration does not alter
the freeze-out results significantly \cite{Betz:2008ka}, we do not include
this effect in the present study for the expanding medium. 
Below, we consider a $5$~GeV trigger parton which corresponds 
to trigger-$p_T$ of $p_T^{\rm trig}=3.5$~GeV assuming that a fragmenting jet creates particles 
with $\sim 70$\% of its energy.
\\\hspace*{4mm}
Experiments can only trigger on the jet direction, thus one has to consider different 
starting points for the jet which is done according to 
$x = r\cos\phi, y = r\sin\phi$, where $r=5$~fm is chosen to model surface emission. 
We incorporate $\phi=90$, $120, 150, 180, 210, 240, 270$~degrees.
\begin{figure}[t]
\centering
\begin{minipage}[c]{4.2cm}
\hspace*{-1.2cm}
  \includegraphics[scale = 0.45]{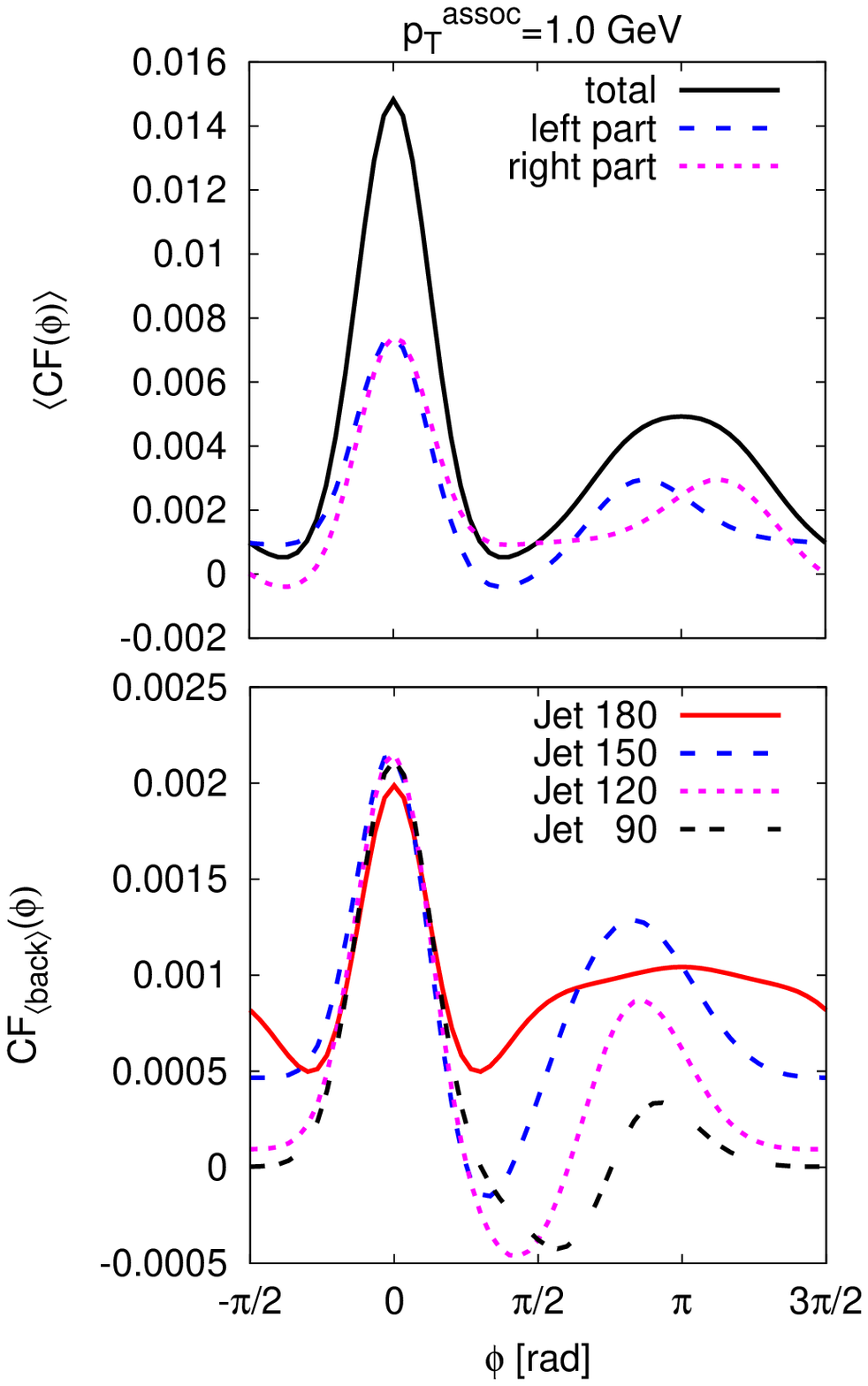}
\end{minipage}
\hspace*{-0.9cm}  
\begin{minipage}[c]{4.2cm} 
  \includegraphics[scale = 0.45]{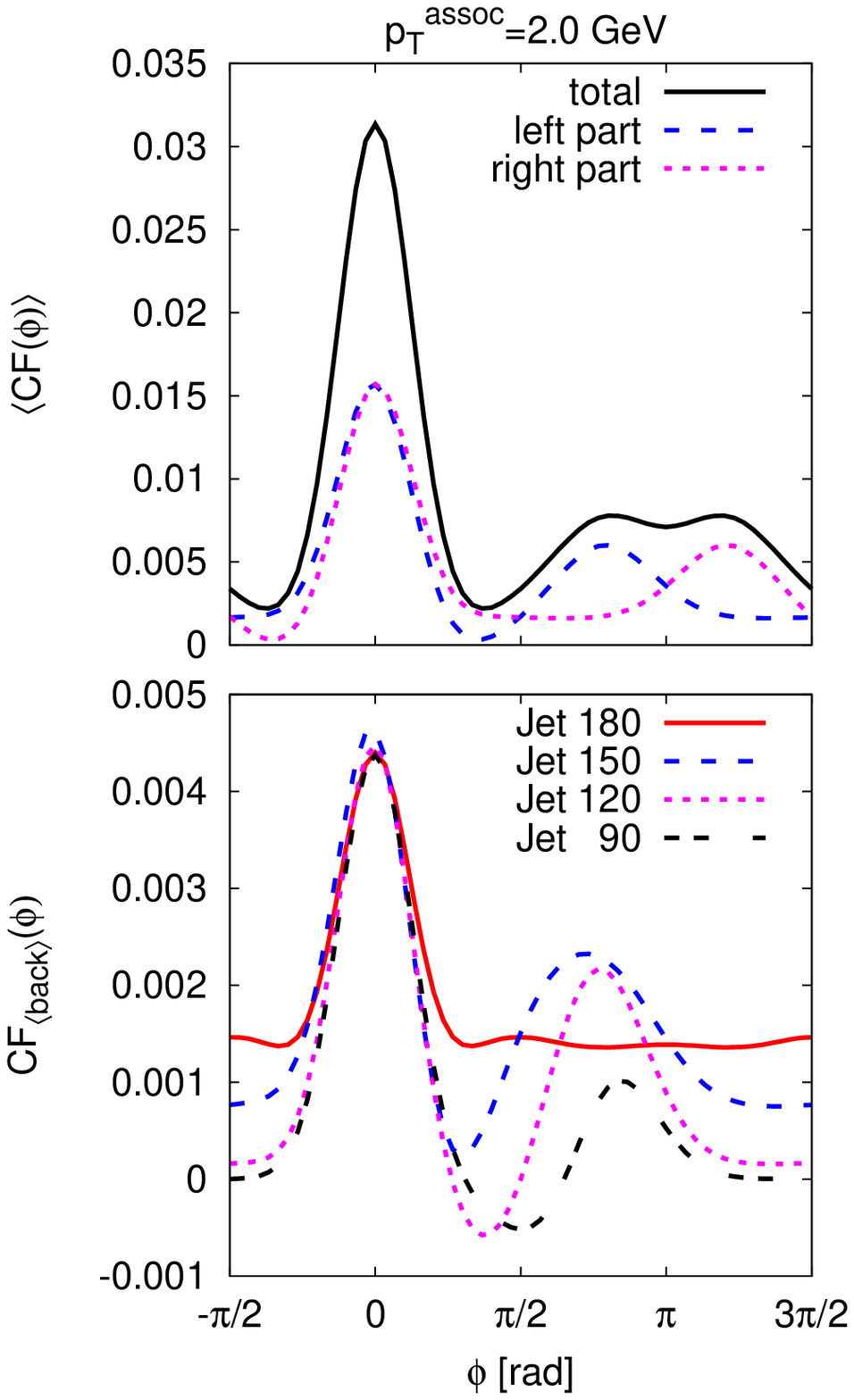}
\end{minipage}
  \caption{The normalized, background-subtracted, and path-averaged azimuthal 
  two-particle correlation after performing an isochronous CF freeze-out 
  (solid lines in the upper panels) for $5$~GeV jets depositing energy and momentum for a 
  $p_T^{\rm assoc}=1$~GeV (left panel) and a $p_T^{\rm assoc}=2$~GeV (right 
  panel). The dashed lines in the upper panels represent the 
  averaged contributions from the different jet paths. The lower 
  panel displays the contribution from the different jet trajectories with
  $\phi=90...180$~degrees.}
  \label{CFIsochronous}\vspace*{-0.5cm}
\end{figure}
To model the experimental situation, the CF freeze-out results are convoluted
by a Gaussian representing the near-side jet, leading to a 
background subtracted, normalized, and jet-averaged CF signal for $b=0$~fm
\begin{eqnarray}\hspace*{-4ex}
\langle CF(\phi)\rangle&=&
\frac{1}{\int_0^{2\pi}\langle N_{\rm back}(\phi)\rangle d\phi}
\left[\frac{d\langle N_{\rm con}\rangle(\phi)}{p_T dp_T dy d\phi}-
\frac{d\langle N_{\rm back}\rangle(\phi)}{p_T dp_T dy d\phi}
\right]\,.
\end{eqnarray}
This CF signal (see solid lines in the upper panels of Fig.\ \ref{CFIsochronous}) 
displays a broad away-side peak for $p_T^{\rm assoc}=1$~GeV, 
while a double-peaked structure occurs for $p_T^{\rm assoc}=2$~GeV. 
The reason is that the contribution of the different paths (for $\phi=90...180$~degrees
see lower panels of Fig.\ \ref{CFIsochronous}) add up to two peaks in the left 
and in the right part of the away-side (dashed lines in the upper panel of Fig.\ \ref{CFIsochronous}).
\\\hspace*{4mm}
It is important to notice that the main contributions to the peaks in the left 
and right part of the away-side come from non-central jets (see lower panel of 
Fig.\ \ref{CFIsochronous}).
\\\hspace*{4mm}
Thus, we have shown, using a full $(3+1)$-dimensional ideal hydrodynamical 
prescription, that a double-peaked away-side structure can be formed due to the 
different contributions of several possible jet trajectories through an 
expanding medium \cite{Neufeld,Chaudhuri}. Therefore, it seems natural 
to conclude that this shape, interpreted as a conical signal, does not result 
from a ``true'' Mach cone, but is actually generated by the averaging of 
distorted wakes. Clearly, the emission angle of such a structure is not 
connected to the EoS. However, these results do not imply that Mach cones 
are not formed in heavy-ion collision. 
The effects of longitudinal expansion, finite impact parameter, and
different freeze-out prescritiptions (like coalescence \cite{GiorgioProceedings})
remain to be considered.



\section*{Acknowledgments} 
The work of B.B.\ is supported by BMBF and by the Helmholtz
Research School H-QM. J.N.\ and M.G.\ 
acknowledge support from DOE under
Grant No.\ DE-FG02-93ER40764. 
G.T.\ was (financially) supported by the Helmholtz International
Center for FAIR within the framework of the LOEWE program
(Landesoffensive zur Entwicklung Wissenschaftlich-\"Okonomischer
Exzellenz) launched by the State of Hesse.

\end{document}